\documentclass[a4paper,12pt,oneside]{article}
\usepackage[warn]{mathtext}
\usepackage{amsmath}
\usepackage[T2A]{fontenc}
\usepackage{latexsym}
\usepackage[cp1251]{inputenc}
\usepackage[english]{babel}
\usepackage{graphics}
\usepackage{indentfirst}
\usepackage{cite}
\usepackage{epsfig}

\title{Quasi-Cherenkov Radiation from Relativistic Particles Passing Through a Photonic Crystal}

\author{V.G. Baryshevsky\thanks{E-mail:bar@inp.bsu.by}, A.A. Gurinovich\thanks{E-mail:gur@inp.bsu.by}}

\begin{document}
         \maketitle
         \begin{center}
                  Research Institute for Nuclear Problems, Belarusian State University,\\
Bobruiskaya 11, 220030 Minsk, Belarus
         \end{center}

\date{}
\topskip 0cm \topmargin -2cm

\headheight 0mm \hoffset -1.0cm \oddsidemargin 2cm

\textwidth 14.0cm \textheight 25.7cm

\begin{abstract}
The expressions for spectral-angular distribution of
quasi-Cherenkov radiation emitted by a relativistic particle
traversing a photonic crystal are derived. It is shown that for a
relativistic particle, passing through a photonic crystal formed
by periodically strained threads, the intensity of quasi-Cherenkov
radiation emitted at small angles to the direction of particle
motion, as contrasted to ordinary Cherenkov radiation, exhibits
anisotropic properties as the the photon momentum is rotated about
the direction of particle motion (as the crystal is rotated about
the direction of particle motion at fixed-angle observation of the
outcoming photon).

The intensity of quasi-Cherenkov radiation in terahertz and
optical ranges is shown to be high enough to allow the
experimental study of quasi-Cherenkov radiation in these frequency
ranges.
\end{abstract}

\section{Introduction}
\label{sec:intro}

Diffraction radiation of photons from relativistic particles
moving in crystals (natural or artificial spatially periodic
structures) has come under intensive theoretical and experimental
investigation in recent years.

The ever-growing interest to the study of radiation and emission
processes in spatially periodic structures was inspired by pioneer
research into
 microwave oscillators using radiation from an electron
 beam in a periodic slow-wave structure (traveling wave tube,
 backward wave oscillator, etc.) \cite{TWT1,TWT2}.

In 1953, Smith and Purcell \cite{SP}  made the next step and
observed generation of incoherent radiation at visible wavelengths
by using a finely-focused electron beam propagating close to the
surface of a metal diffraction grating (at the distance $\delta
\le \frac{{\lambda \beta \gamma} }{{4\pi} }$, $\delta $ is the {
beam impact parameter, $\lambda $ is the radiation wavelength,}
$\beta = v /c$ , $v$ is the electron beam velocity, and $\gamma $
is the electron Lorentz factor).

The Smith-Purcell effect belongs to a general class of diffraction
radiation effects induced by the interaction of electrons with a
medium.
Diffraction of waves associated with the electromagnetic field of
the electron by an obstacle leads to the so-called diffraction
radiation \cite{Bolotovskii}.

The feature of diffraction radiation is  that the radiation
frequency of photons emitted forward in the direction of particle
motion increases as the particle energy is increased, passing from
optical to X-ray and $\gamma$ ranges \cite{Ter-Mikaelyan}. (In
X-ray and $\gamma$ ranges, this type of radiation is called the
resonant radiation.)
In  1971, it was shown \cite{VG1971} that for particles moving in
a crystal, the induced Vavilov-Cherenkov effect arises in the
X-ray range, and thus the spontaneous effect arises too
\cite{PXRbook,VG_NO2012,vg+fer-72}.
The feature of this effect is that the X-ray-quantum  frequency is
independent of the particle energy even for radiation in the
forward direction.
Later this radiation was called the parametric (quasi-Cherenkov)
X-ray radiation (PXR) and  observed experimentally in 1985
\cite{PXRbook,VG_NO2012,VG+Fer,pisma85}.
This effect can arise because the Bragg diffraction of X-ray
photons in crystals has a result that even in the X-ray range,
where the refractive index of a medium $n(\omega) <1$, there are
waves for which the refractive index $n(\omega)$ can be greater
than unity.
A similar phenomenon also occurs in artificial
three(two)-dimensional periodic structures, now frequently
referred to as photonic (electromagnetic) crystals
\cite{nim06}. 
The ''grid'' structure formed by periodically strained dielectric
threads was experimentally studied in \cite{VolumeGrating}, where
it was shown that ''grid'' photonic crystals can have a
sufficiently high $Q$ factor ($10^4 - 10^8$).
Induced radiation from an electron beam in a photonic crystal
formed by periodically strained metallic threads is described
in~\cite{nim06}. 
Radiation generators using two- or three-dimensional periodic
structures (natural, photonic, electromagnetic crystals) are
called volume free electron lasers (VFEL).

A significant number of theoretical and experimental studies of
parametric (quasi-Cherenkov) X-ray radiation are presently
available \cite{PXRbook,VG_NO2012,artru}.

 Behind a theoretical description of parametric X-ray radiation  is  the application of the dynamical theory of diffraction of
 X-ray radiation,  developed in works on Bragg diffraction
 of X-ray photons \cite{Ewald,nim06_James,nan4,4,132}.  The backbone that underlies the dynamical theory of diffraction is
 the possibility to use
 low-order perturbation theory to describe scattering of X-ray quanta by
 atoms.

Typically, in optical and microwave ranges, the perturbation
theory does not apply to describe photon scattering by scatterers
that form an electromagnetic (photonic) crystal as this theory
does in the X-ray range. Nevertheless, as we have demonstrated
earlier \cite{nim06}, using  the scattering amplitude to describe
the interaction of photons with a scattering center we can derive
equations defining dynamical diffraction in electromagnetic
crystals, and so we can use many of the  results obtained in the
PXR theory \cite{nim06,abg_fanem,bunch_arx} to describe the
emission of photons from relativistic particles moving in
electromagnetic (photonic) crystals.

This paper derives the expressions for spectral-angular
distribution of quasi-Cherenkov radiation emitted by a
relativistic particle traversing a photonic crystal. It is shown
that when a  relativistic particle moves in a photonic crystal
formed by periodically strained threads, the intensity of
Cherenkov radiation emitted at small angles to the direction of
particle motion, as contrasted to ordinary Cherenkov radiation,
exhibits anisotropic properties as the photon momentum is rotated
about the direction of particle motion (as the crystal is rotated
about the direction of particle motion at fixed-angle observation
of the outcoming photon).

It is shown that the intensity of quasi-Cherenkov radiation in
terahertz and optical ranges is high enough to allow the
experimental study  of quasi-Cherenkov radiation.

\section{ Spectral-angular distribution of radiation produced by a particle passing through a photonic crystal}
\label{sec:spectral-angular}

 To describe the process of photon emission by a particle traversing matter, we shall use the method developed in
  \cite{PXRbook,VG_NO2012}.

Knowing the field $\vec{E}(\vec{r},\omega)$, produced by a
particle at a large distance from the target, we can find both the
spectral-angular density of radiant energy per unit solid angle,
$W_{\vec{n}\omega}$, and the differential number of emitted
photons,
  $dN_{\vec{n}\omega}=\frac{W_{\vec{n}\omega}}{\hbar\omega}$,
\begin{equation}
\label{berk_2.1}
W_{\vec{n}\omega}=\frac{er^2}{4\pi^2}\overline{\left|\vec{E}(\vec{r},\omega)\right|^2},
\end{equation}
The vinculum here means averaging over all possible states of the
radiating system.

To find the field $\vec{E}(\vec{r},\omega)$, we need to solve
Maxwell's equations describing the interaction of particles with a
medium.
The transverse solution can be found using the Green function of
these equations, which satisfies the expression
\cite{PXRbook,VG_NO2012,lanl_7a}:
\begin{equation}
\label{berk_2.2} G=G_0+G_0\frac{\omega^2}{4\pi
c^2}(\hat{\varepsilon}-1)G.
\end{equation}
Here $G_0$ is the transverse Green function of Maxwell's equation
at $\hat{\varepsilon}=1$ (it is given, for example, in \cite{66})
and $\hat{\varepsilon}$ is the permittivity tensor of the medium.

Using $G$, we can find the field we are concerned with
\begin{equation}
\label{berk_2.3} E_n(\vec{r},\omega)=\int
G_{nl}(\vec{r},\vec{r}^{~\prime},\omega)\frac{i\omega}{c^2}j_{0l}(\vec{r},\omega)d^3
r^{\prime},
\end{equation}
 where $n,\, l =x,\, y,\, z$ and  $j_{0l}(\vec{r},\omega)$ is the Fourier
transformation of the {$l-th$} component of the current produced
by moving charged particles. In the linear approximation, the
current is determined by the particle's  velocity and trajectory,
which can be found by solving the equation of particle motion.
Under quantum-mechanical consideration, the current $j_0$ is the
current of transition of the particle-medium system  from one
state to another.

According to \cite{PXRbook,VG_NO2012},  the Green's function at
$r\rightarrow \infty$ is expressed  through the solution of
homogeneous Maxwell's equations $E_{\vec
kl}^{(-)s}(\vec{r},\omega)$ containing at infinity
 a  converging spherical wave:
\begin{equation}
\label{berk_2.4} \lim_{r\rightarrow \infty}
G_{nl}(\vec{r},\vec{r}^{~\prime},\omega)=
\frac{e^{ikr}}{r}\sum\limits_s e^s_n E^{(-)s*}_{\vec{k}l}(\vec{r}^{~\prime},\omega),\\
\nonumber
\end{equation}
where $\vec k= k\frac{\vec r}{r}$, $\vec{e}^{\,s}$  is the unit
polarization vector, $s=1,2$, and
$\vec{e}^{\,1}\perp\vec{e}^{\,\,2}\perp\vec{k}$.
%
%
%
%

The solution  $E^{(-)s*}_{\vec{k}l}(\vec{r}^{~\prime})$ for large
distances from the area occupied by the medium has the form
\[
\vec{E}_k^{(-)s}(\vec{r}^{~\prime}, \omega)=\vec{e}^{\,\,s}
e^{i\vec{k}\vec{r}^{\,\, \prime}}+\mbox{const}\frac{e^{-i {k}
{r}^{\,\, \prime}}}{r^{\, \prime}}.
\]
The solution $\vec{E}_{\vec{k}}^{(-)s} (\vec{r},\omega)$ is
associated with the solution of homogeneous Maxwell's equations
$\vec{E}_{\vec{k}}^{(+)s}(\vec{r}, \omega)$ that describes a
plane-wave scattering by a crystal and contains at infinity a
 diverging spherical wave
$\vec{E}_k^{(+)s}(\vec{r}, \omega)=\vec{e}^{\, \, s}
e^{i\vec{k}\vec{r}}+\mbox{const}\frac{e^{i {k} {r}}}{r}$, namely,
(see \cite{PXRbook,VG_NO2012}):
\begin{equation}
\label{berk_2.5}
\vec{E}^{(-)s*}_{\vec{k}}=\vec{E}^{(+)s}_{-\vec{k}}.
\end{equation}

Using (\ref{berk_2.3}) and  (\ref{berk_2.5}), we obtain
\begin{equation}
\label{berk_2.6} \vec{E}_n(\vec{r},
\omega)=\frac{e^{ikr}}{r}\frac{i\omega}{c^2} \sum\limits_S
\vec{e}^{\,\,s}_n\int \vec{E}^{(-)s*}_{\vec{k}}(\vec{r},
\omega)\vec{j}_0(\vec{r}^{\, \, \prime}, \omega)d^3 r^{\prime}.
\end{equation}
As a result, the spectral density of radiant energy for photons
with polarization $s$ can be written in the form:
\begin{equation}
\label{berk_2.7}
W_{\vec{n},\omega}^s=\frac{\omega^2}{4\pi^2c^2}\overline{\left|\int\vec{E}^{(-)s*}_{\vec{k}}(\vec{r},
\omega) \vec{j}_0(\vec{r}, \omega)d^3 r\right|^2},
\end{equation}
\begin{equation}
\label{berk_2.8} \vec{j}_0(\vec{r}, \omega)=\int e^{i\omega
t}\vec{j}_0(\vec{r}, \omega) dt=eQ \int e^{i\omega
t}\vec{v}(t)\delta(\vec{r}-\vec{r}(t))dt,
\end{equation}
where $eQ$ is the charge of the particle, $\vec{v}(t)$ and
$\vec{r}(t)$ are the velocity and the trajectory of the particle
at time $t$.

Substitution of (\ref{berk_2.8}) into (\ref{berk_2.7}) gives
\begin{equation}
\label{berk_2.9} dN^s_{\vec{n}, \omega}=\frac{e^2
Q^2\omega}{4\pi^2 \hbar c^3}\overline{\left|\int
\vec{E}^{(-)s*}_{\vec{k}}(\vec{r}(t), \omega)\vec{v}(t) e^{i\omega
t}\right|^2} dt.
\end{equation}
 Integration in (\ref{berk_2.9}) is performed over the entire
domain of particle motion.
Solving  homogeneous Maxwell's equations
 instead of inhomogeneous significantly simplifies the
analysis of the radiation problem and enables considering
different cases of radiation with due account of  multiple
 scattering.


We shall further concern ourselves with the generation of
parametric quasi-Cherenkov radiation by a relativistic charged
particle moving at constant velocity ($\vec v= $const) through a
photonic crystal, i.e. [see (\ref{berk_2.5})],
\begin{equation}
\label{berk_2.9a} dN^s_{\vec{n}, \omega}=\frac{e^2
Q^2\omega}{4\pi^2 \hbar c^3}\overline{\left|\int
\vec{E}^{(+)s}_{-\vec{k}}(\vec{r}(t), \omega)\vec{v}(t) e^{i\omega
t}\right|^2} dt.
\end{equation}
According to (\ref{berk_2.9}), to find the number of photons
emitted by a particle traversing a crystal plate, we need to find
the explicit expressions for the solutions
$\vec{E}^{(-)s}_{\vec{k}}$. To do this, we only need to know the
solutions $\vec{E}^{(+)s}_{\vec{k}}$ describing refraction and
diffraction of photons in the crystal, since
\[
\vec{E}^{(-)s}_{\vec{k}}=(\vec{E}^{(+)s}_{-\vec{k}})^*,
\]
as we have stated  earlier in this section.

\section{Refraction and diffraction of electromagnetic waves in photonic crystals formed by periodically strained threads}
\label{sec:diffraction}

Refraction and diffraction of waves in natural and artificial
(photonic, electromagnetic) crystals have been intensively studied
since the first publications  on  the dynamical theory of X-ray
diffraction in crystals appeared  \cite{Ewald,Laue}.
The feature of the dynamical theory of X-ray diffraction in
crystals is that it allows  the application of the perturbation
theory to describe  scattering of X-ray quanta by atoms
\cite{Cole,BETE}.

Further analysis has shown that even when the perturbation theory
does not apply to describe scattering by a single center belonging
to a crystal (as,  e.g., in the case of neutrons or slow
electrons), the methods similar to those used in the dynamical
theory of X-ray diffraction work well to describe  diffraction of
radiation in crystals.

According to \cite{BETE}, the direction of scattered waves (formed
through diffraction in  crystals) leaving the plane-parallel plate
is uniquely determined by the incident direction and the momentum
(energy, wavelength) of the incident radiation. Moreover, it is
determined in the same way as in the elementary kinematic
 Laue theory of interference developed for thin plates, for which the  effects of refraction may be neglected: the
 projections onto the crystal surface of the wave vectors of scattered and incident waves differ  by the  projection
 of the reciprocal lattice vector, $\vec{\tau}_{\perp}$, onto the crystal surface. The possible refraction only leads
 to the intensity redistribution
 among the scattered (diffracted) waves.

According to \cite{berk_104,nim06,VG_NO2012}, the dynamical theory
of diffraction enables developing a common approach to the
description of photon emission from relativistic particles in
natural and artificial crystals.

Taking this approach to the analysis of the emission process, we
may use similar formulas, varying only the form of the
coefficients that define frequency-angular and time properties of
the produced radiation.
It should be stated here that in contrast to the  Laue-Ewald
dynamical theory, the perturbation theory does not, as a rule,
apply to describe scattering of electromagnetic waves by an
elementary scatterer (crystal cell) in a photonic crystal.

The theory of electromagnetic-wave propagation in photonic
crystals has been actively studied, see, e.g.,
\cite{yablonovitch,Yablonovitch1,Temkin,Temkin-exp,PC2,Sirigiri,Shapiro,Kilin,Kilin1,Silverin}.
The authors focused on calculating the energy-band structure for
such frequency ranges and parameters of photonic crystals for
which a photon forbidden zone exists (the analogue of the electron
forbidden zone in crystals, i.e., the region of total Bragg
reflection).
The authors of \cite{PC4,Suzuki,Nicorovici}  gave a detailed
analysis of the band structure and the field inside crystals for
the case of two-dimensional crystals formed by periodically
strained metallic threads.

At the same time, as has been shown earlier in section
~\ref{sec:spectral-angular}, to investigate the emission of waves
by particles traversing a photonic (electromagnetic) crystal, we
need to know the expression for the electromagnetic field both
inside and outside the crystal. This means that we need to know
the solutions to Maxwell's equations that describe refraction and
diffraction of waves in a photonic crystal.

In this connection we generalized the dynamical theory of
diffraction in crystals to the case of photonic (electromagnetic)
crystals \cite{nim06}. 
Since under this approach the major parameter  is the amplitude of
wave scattering by a scatterer, the approach becomes applicable to
the description of  refraction and diffraction in crystals built
from the elements (threads, spheres) with arbitrary dielectric
permittivity, in which case the perturbation theory does not apply
to describe scattering by a center.

Because the theory  \cite{nim06} will further be used to describe
quasi-Cherenkov radiation in photonic crystals, let us recall the
main aspects of the theory of refraction and diffraction of
electromagnetic waves in the case of interest --- two-dimensional
periodic electromagnetic crystals built from periodically strained
threads \cite{nim06}.

\begin{figure}[h] \epsfxsize = 6 cm
\centerline{\epsfbox{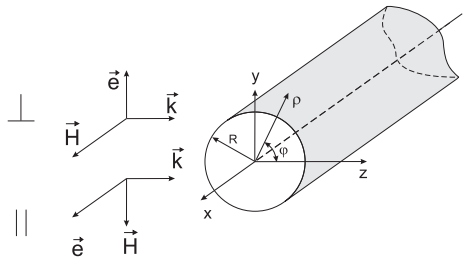}} \caption{Coordinates and
vectors} \label{fig:cylinder}
\end{figure}

Let us suppose that a plane electromagnetic wave characterized by
the electric field vector  $\vec{E} (\vec{r})= \vec{e} e^{i
\vec{k} \vec{r}}$ is incident on the thread.
Here  $\vec{e}$ is the unit polarization vector, $\vec{k}$ is the
wave vector, and  $\vec{r}$ is the radius-vector.
Let us also suppose that the wave is incident perpendicular to the
thread's axis, i.e.,  along the $z$-axis.
The thread is placed at the origin of coordinates and its axis
coincides with the $x$-axis (see Fig. \ref{fig:cylinder}).

Let us assume that the thread is a cylinder of radius $R$.
In this case,  the problem of wave scattering by a cylinder has a
well-known solution (see, e.g., \cite{Nikolsky}).

In particular, if the  polarization vector $\vec{e}$ is parallel
to the thread's axis ($\vec{e} = \vec{e}^{\, \,
\parallel}$), then outside the region occupied by the thread,
i.e., at $\rho > R$, the expression for the wave formed through
scattering can be written as a superposition of the incident
$\vec{E}_0$ and scattered $\vec{E}_{sc}$ waves:
\begin{equation}
\vec{E}=\vec{e} \, \Psi (\rho,z)
\label{eq:111}
\end{equation}

\begin{equation}
\Psi (\rho,z)= e^{i k z} + \sum_{n=-\infty}^{\infty} i^n
a_n^{\parallel} H_n^{(1)} (k \rho),
\label{eq:111.1}
\end{equation}
where $\rho=(y,z)$ is the transverse coordinate, $H_n^{(1)}$ is
the Hankel function of order $n$, and $k=\frac{\omega}{c}$ is the
wave number.

If the dielectric permittivity and magnetic permeability of the
thread are $\varepsilon (\omega)$ and $\mu(\omega)$, respectively,
then the expressions for the amplitudes $a_n$ can be written in
the form
\begin{equation}
a_{n}^{\parallel} = \frac{-J_n(k_t R)
J_n^{\prime}(kR)+\sqrt{\frac{\varepsilon}{\mu}} J_n^{\prime}(k_t
R)J_n(k R)}
{J_n (k_t R) H_n ^{(1)\prime}(kR)-\sqrt{\frac{\varepsilon}{\mu}}
J_n^{\prime}(k_t R)H_n^{(1)}(k R)}~,
 \label{an_full_par}
\end{equation}
where $J_n$ are the Bessel functions of order  $n$,  $k_t= k
\sqrt{\varepsilon (\omega)\mu (\omega)}$; for nonmagnetic metals
$\mu(\omega) = 1$ and $\varepsilon (\omega)=1+ i \frac{4 \pi
\sigma_t}{\omega}$, where $\sigma_t$ is the conductivity of the
thread's material.

Let us now consider scattering by a thread of a wave whose
polarization is orthogonal to the thread's axis,  $\vec{e} \perp
0x$.
It is convenient in this case to consider rescattering of waves
using the amplitude of scattering of a magnetic component of the
electromagnetic wave \cite{VG+Zh_Nanophot}, because the wave's
magnetic field vector $\vec{H}_0=\vec{e}^{\, \,
\parallel} e^{i k z}$  is parallel to the thread's axis.
For the scattered wave, the magnetic field strength  at  $\rho
> R$ has the form
\begin{equation}
\vec{H}=\vec{e}_{\, \, \parallel} \Psi_{H} (\rho,z),~ \Psi_{H}
(\rho,z)= e^{i k z} + \sum_{n=-\infty}^{\infty}i^n a_n^{\perp}
H_n^{(1)} (k \rho) e^{i n \varphi}
\label{Hpar1}
\end{equation}
where $\varphi$ is the azimuth  angle (see Fig.
\ref{fig:cylinder}).
\begin{equation}
a_{n}^{\perp} = \frac{-J_n(k_t R)
J_n^{\prime}(kR)+\sqrt{\frac{\mu}{\varepsilon}} J_n^{\prime}(k_t
R)J_n(k R)}
{J_n (k_t R) H_n ^{(1)\prime}(kR)-\sqrt{\frac{\mu}{\varepsilon}}
J_n^{\prime}(k_t R)H_n^{(1)}(k R)}~,
 \label{an_full_perp}
\end{equation}

Let us suppose that the radiation wavelength $\lambda$ is much
greater than the thread's radius $R$. In this case, analyzing
scattering of a wave whose electric vector $\vec{E}_0$ has a
polarization parallel to the thread's axis, we can retain in the
sum (\ref{eq:111.1}) only the term proportional to
$a_0^{\parallel}$ \cite{nim06}.
Moreover, for well-conducting threads, the amplitude
$a_0^{\parallel} \gg a_n^{\perp}$ \cite{nim06}.
As a consequence, the wave whose polarization is parallel to the
thread is scattered more strongly than the wave whose vector $\vec
E_0$ is polarized orthogonally to the thread
\cite{VG+Fer,pisma85}. Let us also note that according to the
analysis \cite{VG+Zh_Nanophot},
 if the wave's
electric field vector $\vec{E}_0$ is orthogonal to the thread's
axis, then the following relationships hold true for all $|n|>1$:
$a_0^{\perp} \approx a_{\pm 1}^{\perp}$, $a_0^{\perp} \approx
a_0^{\perp}$, and $ a_{\pm 1}^{\perp} \gg a_n^{\perp}$.
This implies that for a wave  with orthogonal polarization, three
terms will retain in the sum (\ref{Hpar1}), as contrasted to
scattering of a wave whose electric field vector $\vec{E}_0$ has a
polarization parallel to the thread's axis.

\subsection{Refraction of waves by a set of scatterers}

Let an electromagnetic wave be incident on a set of threads with
coordinates $\rho_l=(y_l,z_l)$.
For the sake of concreteness, we shall first consider a wave with
polarization parallel to the threads: $\vec{e}=\vec{e}^{\, \,
\parallel}
\parallel 0x$.
The scattered wave can be represented as a superposition of waves
scattered by single threads
\begin{equation}
\Psi=e^{ikz}+a_0^{\parallel} \sum_{l} H_0^{(1)} (k \left\vert
\vec{\rho }-\vec{\rho }_{l}\right\vert)e^{ikz_l}
\label{sum_cylinder}.
\end{equation}
Using the integral representation of the Hankel  function
$H_0^{(1)}(k \rho)$, we can express the scattered wave as
\begin{equation}
\Psi=e^{ikz}+A_0^{\parallel} \sum_{l}
\int\limits_{-\infty }^{\infty }\frac{e^{i k\sqrt{\left\vert \vec{%
\rho }-\vec{\rho }_{l}\right\vert ^{2}-x^{2}}}}{\sqrt{\left\vert
\vec{\rho }-\vec{\rho }_{l}\right\vert ^{2}-x^{2}}}dx e^{ikz_l},
\label{ncylinder}
\end{equation}
where $A_0^{\parallel}=-\frac{i }{\pi} a_0^{\parallel}$ and
$\left\vert \vec{\rho }-\vec{\rho }_{l}\right\vert
^{2}=(y-y_l)^2+(z-z_l)^2$.

Let us consider a wave passing through a layer formed by the
threads whose axes lie in the plane $x0y$ at a distance $d_y$ from
one another.
Let us suppose that the transverse dimension $L_{\perp}$ of the
layer is much greater than the distance $d_y$ and  the radiation
wavelength ($L_{\perp} \gg d_y$ and  $L_{\perp} \gg \lambda$).
Summation over the coordinates  $y_l$ in this case gives the
following expression for   $\Psi$:
\begin{equation}
\Psi=e^{ikz}+\frac{2 \pi i A_0^{\parallel}}{k
d_y}e^{ikz}=(1+\frac{2 \pi i A_0^{\parallel}}{k d_y})e^{ikz}.
\label{average-n-cylinder}
\end{equation}
This expression reflects a well-known fact that a plane wave of
unit amplitude, scattered by a plane layer of scatterers, converts
into a plane wave of complex amplitude
$1+\frac{2 \pi i A_0^{\parallel}}{k d_y}$.
Let us suppose that the parameter $|\frac{2 \pi i
A_0^{\parallel}}{k d_y}| \ll 1$.
In this case, the expression for the wave multiply scattered  by
$m$ number of planes spaced from one another at a distance  $d_z$
can be presented \cite{nim06} in the from $\Psi
=e^{ikn_{\parallel} z}$, where $n_{\parallel}$ is the refractive
index that can be expressed as:
\begin{equation}
n_{\parallel}=1+\frac{2 \pi}{d_y d_z k^2}A_0^{\parallel}.
 \label{n_Bar}
\end{equation}
It follows from (\ref{n_Bar}) that in this case, the real part of
the refractive index $\texttt{Re}\,
n_{\parallel}=n_{\parallel}^{\prime} = 1+ \frac{2 \pi}{d_y d_z
k^2} \texttt{Re}\, A_0^{\parallel}$, while its imaginary part
$\texttt{Im}\, n_{\parallel}=n_{\parallel}^{\prime \prime} =
\frac{2 \pi}{d_y d_z k^2} \texttt{Im}\, A_0^{\parallel}$.

The expressions of the form (\ref{n_Bar}) for the refractive index
can also be derived for a medium made up of wires chaotically
distributed in the plane $(z0y)$.
In this case, the quantity $\frac{1}{d_y d_z}$ is replaced by
$\rho_{yz}$, the density of scatterers in the plane $(z0y)$
($\frac{1}{d_y}\longrightarrow \rho_{y}$, the density of threads
along the $y$-axis, and $\frac{1}{d_z}\longrightarrow \rho_{z}$,
the density of threads along the $z$-axis).

If the threads are chaotically distributed along the $y$-axis,
then $\rho_{yz}=\rho_{y} \frac{1}{d_z}$, where $\rho_{y}$ is the
density of threads along the $y$-axis. When the threads are
distributed chaotically  along the $z$-axis and periodically along
the $y$-axis, the density $\rho_{yz}=\rho_{z} \frac{1}{d_y}$,
where $\rho_{z}$ is the density of threads along the $z$-axis.

Thus for a medium built from chaotically distributed threads, we
have
\begin{equation}
n_{\parallel}=1+\frac{2 \pi \rho_{yz}}{k^2}A_0^{\parallel}.
 \label{n_Bar_chaotic}
\end{equation}
Let us recall here that  $A_0^{\parallel}= - \frac{i}{\pi}
a_0^{\parallel}$.

 A wave with polarization
orthogonal to the thread's axis ($\vec{e} \perp 0x$) can be
considered along similar lines.

The expression for the refractive index $n_{\perp}$ is similar to
(\ref{n_Bar_chaotic}) with   $A_0^{\parallel}$ replaced by
$A_0^{\perp}=-\frac{i}{\pi}(a_0^{\perp}+2 a_1^{\perp})$.
As a result, we have
\begin{equation}
n_{\perp}=1+\frac{2 \pi \rho_{yz}}{k^2}
(-\frac{i}{\pi}a_0^{\perp}-\frac{2 i}{\pi} a_1^{\perp})
\label{n_Bar_perp}
\end{equation}


It is noteworthy that the imaginary part of the refractive index
(\ref{n_Bar}) and (\ref{n_Bar_chaotic}) is nonzero when the wave
is scattered by a perfectly conducting thread when absorption in
the thread is absent.  The result (\ref{n_Bar_chaotic}),
describing the wave passage through a medium formed by chaotically
distributed scatterers, is valid because  the waves scattered at
nonzero angles have random phases, since the threads are
distributed chaotically, and  the scattered waves do not
interfere. As a result, they lead to attenuation of the initial
wave.

In the case (\ref{n_Bar}), the imaginary part must be missing,
because in a crystalline medium scattering of waves is nonrandom.
The attenuation of waves in crystals is determined by their
absorption in the threads \cite{nim06} (see Sec. 3.2).

As the radiation frequency increases and the wavelength becomes
comparable with the radius of the thread ($kR \geq 1$),  other
terms in (\ref{eq:111.1}) and (\ref{Hpar1}) also make
contributions to scattering of waves by a thread.
To find the refractive index in this case, we can also perform
averaging over the location(s) of the threads in the plane
transverse with respect to the $z$-axis, but this procedure
involves cumbersome calculations: first you need to express the
Hankel functions of an arbitrary index $n$ in terms of $H_0$, and
then the problem is reduced to the problem considered earlier in
this section.

Interestingly, in the case when the parameter $k d_z \geq 1$,
i.e., the radiation wavelength is less than the inter-thread
distance along the $z$-axis, the refractive index can be found in
a more simple way.
To do this, we make use of the asymptotic representation of the
Hankel function for  $k \rho \gg 1$
\begin{equation}
\label{eq:22h}
 H_n^{(1)} (k \rho) \approx \sqrt{\frac{2}{\pi k
\rho} }~ e^{i(k \rho -\frac{\pi}{2} (n+ \frac{1}{2}))} = (-i)^n
H_0^{(1)} (k \rho)
\end{equation}
As a consequence,  when the distance from the thread is much
greater than the wavelength ($\rho \gg \frac{1}{k}$), we obtain
the following expression for the scattered wave having the
polarization $\vec{e} = \vec{e}_{\parallel} ~\parallel ~ 0x$
\begin{equation}
\label{eq21}
 \vec{E} = \vec{e}_{\parallel} F^{\parallel}(\varphi)
\sqrt{\frac{2}{\pi k \rho} } e^{i k \rho} e^{-i \frac{\pi}{4}} =
\vec{e} F^{\parallel}(\varphi) H_0^{(1)} (k \rho)
\end{equation}
\begin{equation}
F^{\parallel}(\varphi) = \sum_{n=-\infty}^{\infty} a_n^{\parallel}
e^{i n \varphi}.
\end{equation}
Similar expressions can be written for a wave having the
  electric  vector $\vec{e}=
\vec{e}_{\perp} \perp 0x$ (the magnetic field is directed along
 $0x$).
\begin{equation}
\vec{H} = \vec{e}_{\perp} \sum_{n=-\infty}^{\infty} i^n
a^{\perp}_n
 H_n (k \rho)
F^{\perp} (\varphi) \sqrt{\frac{2}{\pi k \rho} } e^{i k \rho}
e^{-i \frac{\pi}{4}} = \vec{e} F^{\perp}(\varphi) H_0^{(1)} (k
\rho) \label{eq:222}
\end{equation}
\begin{equation}
F^{\perp} (\varphi) = \sum_{n=-\infty}^{\infty} a_n^{\perp} e^{i n
\varphi}
\end{equation}

Let us consider a wave passing through the layer of $N$ number of
scatterers in the plane $z=z_l$.
The electric field strength at point  $z$ is described by the
expression:
\begin{equation}
\vec{E}(\vec{r}) = \vec{e} \sum_{l=1}^{N}
\sum_{n=-\infty}^{\infty} \left( a_n^{\parallel} e^{(i n
\varphi_l)} H_0^{(1)} (k |\vec{\rho} -\vec{\rho_l}|) \right) e^{i
k z_l}, \label{eq:E}
\end{equation}
\begin{equation}
|\vec{\rho} -\vec{\rho_l}|^2=(y-y_l)^2 +(z-z_l)^2, \label{eq:rho}
\end{equation}
where $\varphi_l$ is the azimuth angle between $\vec{k}$ and the
direction  $(\vec{\rho} -\vec{\rho_l})$.

Upon summation over the positions of the threads in the plane
$x0y$,
 using the integration representation of the Hankel
functions
\begin{equation}
H_0^{(1)} (k |\vec{\rho} -\vec{\rho_l}|)= - \frac{i}{\pi}
\int_{-\infty}^{\infty} \frac{e^{i k \sqrt{(\vec{\rho}
-\vec{\rho_l})^2+x^2}}}{\sqrt{(\vec{\rho} -\vec{\rho_l})^2+x^2}}
dx, \label{eq:Hankel}
\end{equation}
and the relations
\begin{equation}
\cos \varphi_l = \frac{z-z_l}{(y-y_l)^2 +(z-z_l)^2}, ~
\sin \varphi_l = \frac{y-y_l}{(y-y_l)^2 +(z-z_l)^2},
\label{eq:cos+sin}
\end{equation}
we obtain the integral expression, which, through integration by
parts, can be reduced to the form ($z > z_l$):
\begin{equation}
\Psi = e^{ikz}+\frac{2 \pi i}{k y} A^{\parallel}(0)+  \text{the
terms}\quad \sim \frac{1}{k(z-z_l)},
\end{equation}
where
\begin{equation}
A^{\parallel}(0)=-\frac{i}{\pi} F^{\parallel}(0) =\left( -
\frac{i}{\pi} \right) \sum_{n=-\infty}^{\infty} a_n^{\parallel}
\label{eq:2222}
\end{equation}

Summation of the waves that have traversed several planes gives
the expression for the refractive index, similar to
(\ref{n_Bar_chaotic}) with $A_0^{\parallel}$ replaced by
$A^{\parallel}(0)$
\begin{equation}
\label{eq:ins1} n_{\parallel}=1+\frac{2\pi\rho_{yz}}{k^2}
A^{\parallel}(0).
\end{equation}

Similarly, for a wave with polarization $\vec{e}_{\perp}$, the
summation of waves (\ref{eq:222}) gives the refractive index of
the form (\ref{eq:ins1}) with $A^{\parallel}(0)$ replaced by
$A^{\perp}(0)$:
\begin{equation}
\label{eq:33.1}
 n_{\perp}= 1 + \frac{2\pi\rho_{yz}}{k^2}
A^{\perp}(0),
\end{equation}
where
$A^{\perp}(0)=-\frac{i}{\pi}F^{\perp}(0)=\left(-\frac{i}{\pi}\right)\sum\limits_{m}a^{\perp}_m$.

 The
wave propagating through 'matter'  built from metallic threads can
be written in the form
\begin{equation}
 \vec E=E_{0\,\parallel}\,\vec
e_{\parallel}\,e^{ikn_{\parallel}z}+ E_{0\perp}\,\vec e_{\perp}\,
e^{ikn_{\perp} z}.
 \label{eq:ins2}
\end{equation}
Since $n_{\parallel}\neq n_{\perp}$, such 'matter'  is
birefringent \cite{VG+Fer,pisma85}, i.e., as the wave propagates,
its linear polarization converts into circular and so on. Because
$\texttt{Im}n_{\parallel} \neq \texttt{Im}n_{\perp}$, such
'matter' exhibits dichroism, and the law of absorption  has a
conventional form  for the polarization $e_{\parallel}(e_{\perp})$
\begin{equation}
\label{ins3} |E_{\parallel(\perp)}|^2 =
|E_{0\parallel(\perp)}|^2\, e^{-2kn_{\parallel(\perp)} z}
\end{equation}
On the other hand, upon introducing the total scattering cross
section by a thread ($\sigma_{\parallel(\perp)}$), we can write
the law of absorption in the form
\begin{equation}
\label{ins4} |E_{\parallel(\perp)}|^2= |E_{0\parallel(\perp)}|^2
e^{-\rho_{ez}\sigma_{\parallel(\perp)}z}.
\end{equation}
From this follows that
\begin{equation}
\label{ins5}
\sigma_{\parallel(\perp)}=\frac{4\pi}{k}\texttt{Im}\,A_{\parallel(\perp)}(0)=-\frac{4}{k}\texttt{Re}\,F_{\parallel(\perp)}(0)
=\frac{4}{k} \sum\limits_n |a_{n_{\perp(\parallel)}}|^2.
\end{equation}
Let us note that the cross section dimension in a 2D case is not a
squared length as in a 3D case, but a length. The amplitude $A(0)$
is a non-dimensional quantity.

Equation (\ref{ins5}) can also be derived from (\ref{eq21}). In a
two-dimensional case, the amplitude $f$ of scattering at the angle
$\varphi$ is defined as the  factor of a cylindrical wave,
$\frac{e^{ik\rho}}{\sqrt{\rho}}: f(\varphi)=\sqrt{\frac{2}{\pi
k}}e^{-i\frac{\pi}{4}} F(\varphi)$.
Let us note that this amplitude has the dimension $\sqrt{[l]}$,
the square root of the length. The amplitude $f(\varphi)$ allows
finding the differential scattering cross section that is defined
by the ratio of the energy flux density in the wave scattered at
the angle $\varphi$ to that in the incident wave
\begin{equation}
\label{ins6} \frac{d\sigma}{d\varphi}=|f(\varphi)|^2=\frac{2}{\pi
k} |F(\varphi)|^2
\end{equation}
From this we immediately obtain:
\begin{equation}
\label{ins7}
\sigma=\int\frac{d\sigma}{d\varphi}d\varphi=\frac{4}{k}\sum\limits_n|a_n|^2.
\end{equation}

Now let us consider a wave reflected in a backward direction from
the layer of cylinders  distributed in the plane $z=z_l$.

In this case, (\ref{eq:E})-(\ref{eq:cos+sin}) yield the following
expression for the wave reflected from the plane and having the
electric field vector parallel to the $x$-axis:
\begin{equation}
E^{\parallel}_{ref}(z<z_l)= \frac{2 \pi i \rho_y}{k} A^{\parallel}
(\pi) e^{2 i k z_l} e^{-i k z}
\label{eq:Ex-reflected}
\end{equation}
\begin{equation}
 A^{\parallel}
(\pi) = - \frac{i}{\pi} F^{\parallel} (\pi) = - \frac{i}{\pi}
\sum_{n=-\infty}^{\infty} (-1)^n a_n^{\parallel}
\label{eq:A(pi)}
\end{equation}
Then we have the following expression for a wave reflected from
$N$ number of planes:
\begin{equation}
E^{\parallel}_{ref}(z)= \frac{2 \pi i \rho_y}{k} A^{\parallel}
(\pi) \sum_{l=1}^N e^{2 i k z_l} e^{-i k z}
\label{eq:Ex-reflectedN}
\end{equation}

When the layer thickness is much greater than the wavelength, we
can perform integration instead of summation and thus obtain the
following expression for  $E^{\parallel}_{ref}$
\begin{equation}
E^{\parallel}_{ref}(z)= -\frac{ \pi \rho_{yz}}{k^2} A^{\parallel}
(\pi) e^{-i k z}
\label{eq:Ex-reflectedNint}
\end{equation}

We can see that, as expected, the amplitude of the wave reflected
from the layer is determined by the amplitude $F(\pi)$ of the wave
scattered in the direction of reflection [see (\ref{eq21}) and
(\ref{eq:A(pi)})], i.e., backwards ($F^{\parallel}(\pi)$ -- for a
wave with parallel polarization and $F^{\perp}(\pi)$ -- for a wave
with orthogonal polarization).

At the same time, the refractive index is determined by the
amplitude $F^{\parallel}(0)$ of a wave scattered in the forward
direction [see (\ref{eq:2222}) and (\ref{eq:ins1})].
From this it immediately follows that when scattering is
non-isotropic (versus isotropic), the mirror-reflection
coefficient cannot be found from the refractive index (dielectric
permittivity) \cite{VG_NO2012}.

\subsection{Refraction and diffraction in a photonic crystal}
\label{sec:3.2}

Now let us consider how the expression for the refractive index
changes as the wave moves in the crystal.

According to \cite{nim06}, when performing  summation of scattered
waves in the case of electromagnetic crystals, it is essential
that rescattering of waves be taken into account. This leads to
vanishing of the imaginary part of the refractive index for
crystals formed by perfectly conducting metallic threads (with
chaotically distributed threads, the imaginary part of the
refractive index is nonzero, as shown earlier).

It is also demonstrated in \cite{nim06} that when electromagnetic
crystals are built from periodically strained parallel metallic
threads, taking rescattering into account allows us to obtain the
following expression for the refractive index of a wave whose
electric field vector $\vec{E}_0$ is parallel to the thread's axis
(when the radiation wavelength is much greater than the thread's
radius $\lambda >> R$):
\begin{equation}
{n}_{\parallel}^2=1+\frac{4 \pi }{\Omega_2 k^2}~\frac{A_0}{1+i \pi
A_0-2CA_0} = 1 + \frac{\eta^{\parallel}}{k^2}~, \label{n15_1}
\end{equation}
where
\[
\eta^{\parallel}=\frac{4 \pi }{\Omega_2
}~\frac{A_0^{\parallel}}{1+i \pi
A_0^{\parallel}-2CA_0^{\parallel}}, \]
$\Omega_2$ is the volume of the unit cell of the 2D-crystal,
$C=0.5722$ is the Euler constant, and $A_0^{\parallel}=-\frac{i
a_0^{\parallel}}{\pi}$.

Let us note that when $kR \ll 1$, the term $(2 C A_0^{\parallel})$
is small as compared to unity, and so it can be discarded.

Let us now consider the refractive index $n_{\perp}$ of a wave
whose electromagnetic field vector is orthogonal to the thread's
axis, $\vec{E}_0 \perp 0x$.
As shown in \cite{nim06}, at $\lambda \gg R$, the inequality
$|n_{\perp}^2 -1| \ll |n_{\parallel}^2 -1|$ holds for
well-conducting threads. The explicit expressions for the
refractive index $n_{\perp}$ were derived in \cite{VG+Zh_Nanophot}
[see formulas (\ref{eq49}) and (\ref{eq:n_2}) below].

Consequently, for a photonic crystal of thickness $L$ satisfying
the condition
$\frac{1}{k |n_{\parallel}-1|} \leq L < \frac{1}{k
|n_{\perp}-1|}$,
refraction can be neglected if the wave's   polarization is
$\vec{E}_0 \perp 0x$, but should be considered if its polarization
is $\vec{E}_0
\parallel 0x$

 Note here that according to \cite{VG+Zh_Nanophot}, at $kR
\ll 1$ the refractive index $n_{\perp}$ for well-conducting
metallic threads is greater than unity ($n_{\perp} > 1$), so at
the lengths $L\geq \frac{1}{k|n_{\perp}-1|}$, the refraction of
waves with polarization $\vec E_0 \perp 0x$ becomes important, and
we can observe the Vavilov-Cherenkov effect \cite{VG+Zh_Nanophot}.

Now let us consider the diffraction of waves in electromagnetic
crystals. The theory of wave diffraction by periodically strained
threads in the case when the wavelength is much greater than the
thread's radius ($kR \ll 1$) is given in \cite{nim06}.
According to \cite{nim06}, the effective dielectric permittivity
of a crystal for a wave with $\vec{E}_0
\parallel 0x$ can be written as
\begin{equation}
\varepsilon^{\, \, \parallel} (\vec{r})= \sum_{\vec{\tau}}
\varepsilon_{\tau}^{\parallel} e^{i \vec{\tau} \vec{r}},
\label{eq:epsilon}
\end{equation}
where the Fourier expansion coefficients
$\varepsilon_{\tau}^{\parallel}$  of the function with crystalline
periodicity $\varepsilon^{\, \, \parallel} (\vec{r})$ have the
form
\begin{equation}
\varepsilon_{0}^{\parallel}=n_{\parallel}^2=1+\chi_{\parallel},~
\chi_{\parallel}=\frac{\eta_{\parallel}}{k^2}=\frac{c^2
\eta_{\parallel}}{\omega^2},~
\varepsilon_{\tau}^{\parallel}=\varepsilon_{-\tau}^{\parallel}=\chi_{\parallel}
\nonumber
\end{equation}
where $\chi_{\parallel}$ is the polarizability, $\vec{\tau}$ is
the reciprocal lattice vector of a 2D periodic photonic crystal,
and   $\vec{\tau}=(0,\tau_y,\tau_z)$.

Note that according to the analysis\cite{nim06}, $\eta_{\parallel}
< 0$ for high-conductivity metallic threads.

 As a result, the quantities
$\varepsilon_{\tau}^{\parallel}$ in this case are similar to the
quantities $\varepsilon_{\tau}$ describing the  diffraction of
hard X-rays. Recall that in the X-ray range, $\varepsilon_{\tau}
<0$ for frequencies much greater than characteristic atomic
frequencies, $\varepsilon_0=n^2=1-\frac{\omega_L^2}{\omega^2} <1
$, where $\omega_L$ is the plasma frequency
\cite{4,nan4,nim06_James}.

The results obtained here allow us to apply the elaborate
dynamical theory of X-ray diffraction to the description of  wave
diffraction  and   photon emission by relativistic particles in
photonic crystals.

Let us note that the equations that allow describing diffraction
in electromagnetic crystals in the case of anisotropic scattering
of waves with polarization parallel $\vec{E} \parallel 0x$ and
orthogonal $\vec{E} \perp 0x$ to the thread were derived in
\cite{VG+Zh_Nanophot}.

\section{Quasi-Cherenkov radiation from charged particles in photonic crystals built from periodically strained threads}
 \label{sec:radiation}

Let a relativistic charged particle of charge $ e Q$ ($|e|$ is the
electron charge, $Q$ is the integer) pass through a photonic
crystal.
According to the analysis in Section \ref{sec:3.2}, in the case of
well-conductive threads,  the refractive index for a wave whose
polarization $\vec{E}$ is parallel to the threads is
$n_{\parallel} < 1$. This means that a particle uniformly moving
in the crystals will not emit photons with such polarization
because of the Vavilov-Cherenkov effect.

However, the situation becomes entirely different when considering
that during the emission the photons produced in the crystal can
undergo diffraction.

According to \cite{VG_NO2012}, the diffraction of virtual photons
in crystals can be described with a set of refractive indices
$n_{\mu} (\omega,k)$, some of which may appear greater than unity
even if the refractive index $n$ in the absence of diffraction was
less than unity.
Particularly, when a diffracted wave is formed in a crystal
through Bragg diffraction, along with a wave with wave vector
$\vec{k}$ (two-wave Laue-Bragg diffraction),
 the interaction of the electromagnetic
wave with the crystal is characterized by two refractive indices:
$n_1 (\omega,k) >1$ and $n_2 (\omega,k) <1$.

As a consequence, two types of waves propagating in a photonic
crystal correspond to each polarization: a fast wave ($n_2 <1$)
and a slow one ($n_1
>1$).
The conditions for Cherenkov radiation can be fulfilled for a slow
wave, but not for a fast one.

A fast wave is emitted at the vacuum-crystal boundary in a similar
manner as in the case of ordinary transition radiation. \emph{}
Let a relativistic particle move in a photonic crystal at constant
velocity $\vec{v}$.
It follows from (\ref{berk_2.9}) and (\ref{berk_2.9a}) that
spectral-angular distribution of the number of photons emitted by
a particle moving through the crystal is determined by the
expression of the form:
\begin{equation}
\frac{d^2 N^s}{d \Omega d \omega} = \frac{e^2 Q^2 \omega}{4 \pi^2
\hbar c^3} \left|\int \vec{E}^{s(-) *}_{\vec{k}} (\vec{r}(t),t)
\vec{v} e^{i \omega t} dt \right|^2
\label{eq:spectral}
\end{equation}
where $\vec{k}=\frac{\omega}{c}\vec{n}$, $\vec{n}$ is the unit
vector directed to the point of observation, and  $\vec{E}^{s(-)
*}_{\vec{k}}=\vec{E}^{s(+)}_{-\vec{k}}$.

In view of (\ref{eq:spectral}), to obtain the spectral-angular
distribution, we need to find the solutions that determine the
plane-wave diffraction in a photonic crystal.
 This solution was found in \cite{nim06} for a wave
with electric field vector parallel to the thread's axis $\vec{E}
\parallel 0x$ for $kR \ll 1$, in which case wave scattering by a thread is
isotropic.
In the general case of anisotropic scattering,  the equations
describing the dynamical diffraction of waves with polarizations
parallel $\vec{E}
\parallel 0x$ and  orthogonal $\vec{E} \perp 0x$ to
the thread's axis in photonic crystals were derived in
\cite{VG+Zh_Nanophot}

We shall make use of the results reported in
\cite{nim06,VG+Zh_Nanophot} to find the spectral-angular
distribution of quasi-Cherenkov radiation  in the case when two
coupled waves are produced through diffraction in a photonic
crystal (two-wave diffraction case).  The waves propagating in a
crystal are described by the following  set of equations:
\begin{eqnarray}
\label{eq:spectral1}
 \left(   \frac{k^2 c^2}{\omega^2} -1 -
g^{s*}_0  \right) \vec{E}^{s(-)}_{\vec{k}} - g^{s*}_{-\vec{\tau}}
\vec{E}^{s(-)}_{\vec{k}_{\tau}} = 0 \\ \nonumber
\left(   \frac{k^2 c^2}{\omega^2} -1 - g^{s*}_0  \right)
\vec{E}^{s(-)}_{\vec{k}_{\tau}} - g^{s*}_{\vec{\tau}}
\vec{E}^{s(-)}_{\vec{k}} = 0
\end{eqnarray}
where $\vec{k}_{\tau}=\vec{k}+\vec{\tau}$, $\vec{\tau}$ is the
reciprocal lattice vector, and $s=1,2$ ($1$ and $2$ correspond to
the polarizations  $\vec{E}$ parallel and orthogonal to the
threads, respectively).
\begin{equation}
\label{eq48}
 g^{(1)}_0 \equiv g^{\parallel}_0= \frac{4 \pi
c^2}{\omega^2 \Omega_2} \frac{A^{\parallel}_0}{1 + i \pi
A^{\parallel}_0}, ~ g^{(1)}_{\tau} \equiv
g^{\parallel}_{\tau}=g^{\parallel}_{0},
 \end{equation}

\begin{equation}
\label{eq49}
 g^{(2)}_0 \equiv g^{\perp}_0=\frac{4 \pi
c^2}{\omega^2 \Omega_2} \left[ \frac{A^{\perp}_0}{1 + i \pi
A^{\perp}_0} + \frac{A^{\perp}_1}{1 + i \frac{\pi}{2} A^{\perp}_1}
\right],
\end{equation}

\begin{equation}
\label{eq:spectral2}
 g^{(2)}_{\vec{\tau}} \equiv
g^{\perp}_{\vec{\tau}}= \frac{4 \pi c^2}{\omega^2 \Omega_2} \left[
\frac{A^{\perp}_0}{1 + i \pi A^{\perp}_0} + \frac{A^{\perp}_1}{1 +
i \frac{\pi}{2} A^{\perp}_1} \cos (\vec{k},\vec{k}+\vec{\tau})
\right],
\end{equation}
Because the diffraction process in a periodic medium has a general
character, equations (\ref{eq:spectral1}) are similar to the
equations of the dynamical  theory of diffraction of X-rays and
neutrons in crystals.
Of course, for X-rays or neutrons, the coefficients similar to
$g_{\tau}$, appearing in (\ref{eq:spectral1}), will be different.

 Particularly, given the  threads for which
$|\varepsilon - 1|,|\mu - 1|\gg 1$, the angular dependence of the
coefficient $g^{(2)}_{\tau} = g^{\perp}_{\tau}$, describing
scattering of a wave whose electric-field polarization vector lies
in the diffraction plane $(\vec k, \vec k + \vec \tau)$, is
defined by two terms: independent ($\sim A_0^{\perp}$) and
dependent ($\sim A_1^{\perp}$) of the scattering angle.
For X-rays, the angular dependence of the quantity, analogous to
$g^{\perp}_{\tau}$, is determined alone by the term proportional
to $\cos(\vec k, \vec k + \tau)$.
If the inequality $|\varepsilon - 1|,|\mu - 1|\ll 1$ for
dielectric (magnetic) permeability of the thread holds, then the
terms proportional to $A_0^{\perp}$ in the expression for
$g^{\perp}_{\tau}$ can be neglected (the coefficient
$A_0^{\perp}\ll A_1^{\perp}$ \cite{VG+Zh_Nanophot}), and the
angular dependence $g^{\perp}_{\tau}$ for the thread will appear
similar to the quantity  $g^{\perp}_{\tau}$ for X-rays.

 Let us note here that when a photonic crystal is made
in a homogeneous isotropic crystal (e.g. when the threads are
submerged in a substance whose dielectric permittivity
$\varepsilon_{av}= 1 + \chi_{av}$), the quantity  $\chi_{av}$ must
appear in (\ref{eq48}) and (\ref{eq49}). For the case when
$\chi_{av} \sim 1$ and  $g_{01} \, g_{\tau} \ll \chi_{av}$, the
diffraction of waves in a periodic medium was considered in
\cite{max94}

 The requirement that the  solution to linear system
(\ref{eq:spectral}) exist yields the dispersion equation defining
vectors $\vec{k}$ that may be present in the crystal. These
vectors are convenient to write in the form (compare with
\cite{lanl_7a}).
\[
\vec{k}_{\mu s}=\vec{k}+\vec{\kappa}^*_{\mu s}\vec{N},\qquad
\kappa_{\mu s}^*=\frac{\omega}{c\gamma_0}\varepsilon^*_{\mu s},
\]
where $\mu=1,2$ and  $\vec{N}$ is the unit vector normal to the
crystal surface through which the particle enters
\begin{equation}
\label{para_1.17} \varepsilon_{\mu
s}=\frac{1}{4}\left[(1+\beta_1)g^s_0-\beta_1\alpha_B\right]
\pm\frac{1}{4}\left\{\left[(1-\beta_1)g^s_0+\beta_1\alpha_B\right]^2+
4\beta_1 g^s_{\vec{\tau}}g^s_{\vec{-\tau}}\right\}^{-1/2},
\end{equation}
$\alpha_B=(2\vec{k}\vec{\tau}+\tau^2)k^{-2} $ is detuning from the
Bragg condition ($\alpha_B=0$ if the Bragg condition is exactly
fulfilled)
\[
\gamma_0=\vec{n}_{\gamma}\cdot\vec{N},\quad
\vec{n}_{\gamma}=\frac{\vec{k}}{k},\quad
\beta_1=\frac{\gamma_0}{\gamma_1}, \quad
\gamma_1=\vec{n}_{\gamma\tau}\cdot\vec{N},\quad
\vec{n}_{\gamma\tau}=\frac{\vec{k}+\vec{\tau}}{|\vec{k}+\vec{\tau}|}.
\]
The general solution of (\ref{eq:spectral}) inside the crystal
reads
\begin{equation} \label{para_1.18}
\vec{E}^{(-)s}_{\vec{k}}(\vec{r})=\sum\limits^2_{\mu=1}\left[\vec{e}^{\,
\, s} \Phi^s_{\vec{k}_{\mu}}\exp(i\vec{k}_{\mu s}\vec{r})+
\vec{e}^{ \, \, s}_{\tau}\Phi^s_{\vec{k}_{\mu \vec{\tau}
}}\exp(i\vec{k}_{\mu s\tau}\vec{r})\right].
\end{equation}
The unit vector  $\vec{e}^{(1)} = \vec{e}^{(1)}_{\tau}$  lies in
the plane formed by the thread's axis (x-axis) and the $0z$-axis
that is orthogonal to the thread;
we assume that the angle between $\vec{e}^{(1)}_{\tau}$ and the
$0x$-axis is small.
As the angle between $\vec{e}^{(1)}$ and the $0x$-axis increases,
the solutions come in  a more complicated form, and so we shall
not consider this case here.
Vectors  $\vec{e}^{(2)}$ and $\vec{e}^{(2)}_{\tau}$ are orthogonal
to the thread's axis and lie in the plane formed by vectors
$\vec{k}$ and $\vec{k}+ \vec{\tau}$;
$\vec{e}^{(2)}_{\tau} \perp \vec{k}+ \vec{\tau}$.

Joining this solution with the solutions to Maxwell's equations in
a vacuum, we can find the explicit form of the field
$\vec{E}^{(-)s}_{\vec{k}}(\vec{r})$ throughout the  area.

Thus, in the case of two-wave dynamical diffraction, a photonic
crystal is characterized by two effective refractive indices for
each polarization:
\begin{equation}
n_{\mu s}=1+ \varepsilon_{\mu s}, ~ \mu=1,2 \label{eq:n_1}
\end{equation}

Let us recall here that in the case under consideration
$|\varepsilon_{\mu s}| \ll 1$.

When the parameter $\alpha_B$  increases and becomes much larger
than $g_{\tau}$, the amplitude of the diffracted wave
 diminishes rapidly, and only the initial wave remains whose
 refractive index is
\begin{equation}
n_s=1+ \frac{1}{2} g^s_0 \label{eq:n_2}
\end{equation}

Analyzing the radiation process in crystals,  we should
distinguish between two cases: quasi-Cherenkov radiation during
the diffraction of the emitted photon in Bragg and Laue
geometries.

\textbf{(a)} Let us consider quasi-Cherenkov radiation in the Laue
geometry.
In the Laue geometry, the incident (transmitted) and diffracted
waves leave the crystal through the same surface: $k_z>0,
k_z+\tau_z>0$, where the $z$-axis is parallel to the normal $N$ to
the crystal surface and directed towards the interior of the
crystal.
Joining the solutions to Maxwell's equations on the crystal
surface via (\ref{eq:spectral}), (\ref{para_1.17}), and
(\ref{para_1.18}), we can obtain for the Laue case:
\begin{eqnarray}
\label{para_1.19} & &
\vec{E}^{(-)s}_{\vec{k}}=\left\{\vec{e}^s\left[-\sum_{\mu=1}^2\xi_{\mu
s}^{0*}e^{-i\frac{\omega}{\gamma_0}\varepsilon^*_{\mu
s}L}\right]e^{i\vec{k}\vec{r}}+
e^s_{\vec{\tau}}\beta_1\left[\sum_{\mu=1}^2\xi_{\mu s}^{\tau*}e^{-i\frac{\omega}{\gamma_0}\varepsilon^*_{\mu s}L}\right]e^{i\vec{k}_{\tau}\vec{r}}\right\}\theta(-z)\nonumber\\
& & +\left\{\vec{e}^s\left[-\sum_{\mu=1}^2\xi_{\mu s}^{0*}e^{-i\frac{\omega}{\gamma_0}\varepsilon^*_{\mu s}(L-z)}\right]e^{i\vec{k}\vec{r}}+e^s_{\vec{\tau}}\beta_1\left[\sum_{\mu=1}^2\xi_{\mu s}^{\tau*}e^{-i\frac{\omega}{\gamma_0}\varepsilon^*_{\mu s}(L-z)}\right]e^{i\vec{k}_{\tau}\vec{r}}\right\}\nonumber\\
& & \times \theta(L-z)\theta(z)+\vec{e}^s
e^{i\vec{k}\vec{r}}\theta(z-L),
\end{eqnarray}
where
\begin{eqnarray}
 \xi^0_{1,2 s}=\mp\frac{2\varepsilon_{2,1
s}-g^s_0}{2(\varepsilon_{2s}-\varepsilon_{1s})}; \nonumber
\\ \xi^{\tau}_{1,2
s}=\mp\frac{g^s_{-\tau}}{2(\varepsilon_{2s}-\varepsilon_{1s})};
\nonumber \\
\theta(z)=\left\{ \nonumber
\begin{array}{l}
1, ~~\mbox{if} ~z\geq 0 \\ 0, ~~\mbox{if} ~z<0.
\end{array}
\right.
\end{eqnarray}

 Equation (\ref{para_1.19}) coincides with a similar
expression derived in \cite{lanl_7a} and  goes over to that in
\cite{lanl_7a} if the quantities $\varepsilon_{\mu\, s},\, g_0$,
and $g_{\tau}$ defined by (\ref{eq48}), (\ref{eq49}),
(\ref{eq:spectral2}), and (\ref{para_1.17}) are replaced by
$\varepsilon_{\mu\, s},\, g_0$, and $g_{\tau}$  defined for X-ray
diffraction (see \cite{lanl_7a}).

Upon substituting (\ref{para_1.19}) into (\ref{berk_2.9}), we can
find the differential number of quasi-Cherenkov-radiation quanta
having the polarization $\vec{e}_s$ and emitted in the forward
direction at small angles to the particle velocity  direction in
the Laue geometry
\begin{eqnarray}
\label{para_1.21} & &\frac{d^2N^L_{0s}}{d\omega
d\Omega}=\frac{e^2Q^2\omega}{4\pi^2\hbar c^3}(\vec{e}^s\vec{v})^2
\left|\sum_{\mu=1,2}\xi_{\mu s}^0
e^{i\frac{\omega}{c\gamma_0}\varepsilon_{\mu
s}L}\left[\frac{1}{\omega-\vec{k}\vec{v}}
-\frac{1}{\omega-\vec{k}^*_{\mu s}\vec{v}}\right]\right.\nonumber\\
& &\left.\times[e^{i(\omega-\vec{k}^*_{\mu
s}\vec{v})T}-1]\right|^2,
\end{eqnarray}
where  $T=L/c\gamma_0$ is the particle's  time of flight,
$\vec{e}_1\parallel [\vec{k}\vec{\tau}]$ and
$\vec{e}_2\parallel[\vec{k}\vec{e}_1]$.

It is obvious that (\ref{para_1.21}) is very similar to the
expression for the spectral-angular distribution of Cherenkov and
transition radiations in a medium having the refractive index
$n_{\mu s}=k_{z\mu s}/k_z=1+\kappa_{\mu s}/k_z=1+\varepsilon_{\mu
s}$.

Spectral-angular distribution of photons in the diffraction
direction $\vec{k}_{\tau}=\vec{k}+\vec{\tau}$ can be derived from
(\ref{para_1.21}) by substituting
\begin{eqnarray*}
& &\vec{e}_s\rightarrow\vec{e}_{s\tau},\qquad \xi^0_{\mu s}\rightarrow \beta_1\xi_{\mu s}^{\tau},\\
& &\xi^{\tau}_{1(2)s}=\pm\frac{g^s_{\tau}}{2(\varepsilon_{1s}-\varepsilon_{2s})}\\
& &\vec{k}\rightarrow\vec{k}_{\tau}, \quad\vec{k}_{\mu
s}\rightarrow\vec{k}_{\tau\mu s}=\vec{k}_{\mu s}+{\tau}.
\end{eqnarray*}

\textbf{(b)} Now let us consider quasi-Cherenkov radiation in a
photonic crystal in the Bragg geometry.
In this case, the diffracted wave leaves the crystal through the
same surface on which the incident wave falls.

Joining the solution to Maxwell's equations (\ref{para_1.18}) with
the solution to Maxwell's equations in a vacuum,  we can obtain
the expressions describing the Bragg geometry.

Spectral-angular distribution of photons emitted in the direction
of particle motion can be derived by substituting (compare with
\cite{lanl_7a}) $\xi^0_{\mu s}\rightarrow \gamma_{\mu s}$
\begin{equation}
\label{para_1.22} \gamma^0_{1(2)s}=
 \frac{2\varepsilon_{2(1)s}-g^s_0}{(2\varepsilon_{2(1)s}-g^s_0)
-(2\varepsilon_{1(2)s}-g^s_0)
e^{i\frac{\omega}{\gamma_0}(\varepsilon_{2(1)s}-\varepsilon_{1(2)s})L}}
\end{equation}

Spectral-angular distribution of photons emitted in the
diffraction direction can be obtained from (\ref{para_1.21}) by
substituting
\begin{eqnarray*}
\vec{e}_s\rightarrow\vec{e}_{s\tau},\quad
\vec{k}\rightarrow\vec{k}_{\tau}, \quad k_{\mu
s}\rightarrow\vec{k}_{\mu\tau s}, \quad
\xi^0_{\mu s}e^{i\frac{\omega}{\gamma_0}\varepsilon_{\mu
s}L}\rightarrow \gamma^{\tau}_{\mu s},
\end{eqnarray*}
where
\begin{equation}
\gamma^{\tau}_{1(2)s}= - \frac{\beta_1 g^s_{\tau}}
{(2\varepsilon_{2(1)s}-g^s_0)-(2\varepsilon_{1(2)s}-g^s_0)e^{i\frac{\omega}{\gamma_0}(\varepsilon_{2(1)s}-\varepsilon_{1(2)s})L}}.
\end{equation}

The formulas derived here describe quasi-Cherenkov radiation in
photonic crystals formed by periodically strained threads, as well
as transition radiation and Cherenkov radiation that is possible
for a state with polarization orthogonal to the threads far from
the region of diffraction reflection.

Let us analyze the expressions derived here.

We shall now consider quasi-Cherenkov radiation in a photonic
(electromagnetic) crystal built from  high-conductivity metallic
threads under the conditions when $kR=\frac{2 \pi R}{\lambda} \ll
1$ ($\lambda$ is the radiation wavelength).
In this case, $\varepsilon^{\perp}_{1,2} \ll
\varepsilon^{\parallel}_{1,2}$, as stated earlier.
Because the number of quanta emitted at the diffraction peak is
proportional to $\xi^{\tau}_{\mu s} \sim g^s_{\tau}$, the number
of  quasi-Cherenkov radiation quanta $N_{\perp}$ with polarization
orthogonal to the thread's  axis is much less than
 the number of quasi-Cherenkov radiation quanta $N_{\parallel}$
with electric-field-vector polarization in the plane formed by the
thread's axis and the wave vector $\vec{k}$ of the emitted wave.
As a result, the polarization of the emitted photon lies mainly in
the plane formed by the thread's axis and the photon's wave
vector.

Let now a particle move along the $z$-axis, in the direction
orthogonal to the entrance surface of the crystal.

We shall focus on the angular distribution of quasi-Cherenkov
radiation emitted either in the forward direction at small angles
with respect to  particle velocity or in the backward direction at
small angles with respect to the direction ($-\vec{v}$) (backward
Bragg reflection).
In this case, the radiation intensity of photons whose
polarization $\vec{e} \parallel 0x$  equals zero if the photon
wave vector $\vec{k}$ lies in the plane $(y,z)$,
and only a significantly less intense radiation remains, whose
polarization is $\vec{e} \perp 0x$.

If we rotate vector $\vec{k}$ about $\vec{v}$ (about the
$z$-axis),  for vector $\vec{k}$ lying  in the $({x},{z})$ plane,
the radiation intensity of photons with polarization $\vec{e}$ in
the $(x,z)$ plane attains its maximum. As is seen, the radiation
intensity exhibits axial asymmetry.

This absence of the azimuth symmetry in the angular distribution
of quasi-Cherenkov radiation in the considered photonic crystal is
in sharp contrast to the case of  ordinary Cherenkov radiation or
parametric X-ray radiation emitted in the forward direction, of
which the azimuth symmetry of angular distribution with respect to
particle velocity is typical.

Let us estimate the number of quasi-Cherenkov quanta emitted by a
particle after it has passed through the crystal of thickness $L$.

It follows from (\ref{para_1.21}) (for detail see
\cite{PXRbook,VGChannelling})  that the number of the emitted
quanta  can be approximately estimated by a simple formula
\begin{equation}
N_{ph} \approx \frac{\alpha Q^2 \left| g^{\parallel}_{\tau}
(\omega_B)\right|^2 \frac{\omega_B}{c} L}{\sin^2 \theta_B  }.
\label{eq:number_ph}
\end{equation}
i.e., in the case of interest, when  $N^{\parallel}$, we have
\begin{equation}
N^{\parallel}_{ph} \approx \frac{\alpha Q^2}{\sin^2 \theta_B}
\left| \frac{4 \pi c^2}{\omega^2_B \Omega_2}
\frac{A^{\parallel}_{0}}{1+ i \pi A^{\parallel}_{0}} \right|^2 k_B
L,
\label{eq:number_ph_par}
\end{equation}
where  $\alpha$ is the fine structure constant.

Recall here that the Bragg diffraction conditions can be written
as

\begin{equation}
m \lambda_B = 2 d {\sin \theta_B},
\label{eq:bragg}
\end{equation}
where $d$ is the lattice  spacing.
Thus,  we can readily obtain for $\omega_B$ and $k_B$
\begin{equation}
\omega_B= m \frac{\pi c}{d \sin \theta_B} = \frac{\tau c}{2 \sin
\theta_B}, ~ k_B= \frac{\tau}{2 \sin \theta_B}
\label{eq:omegaB}
\end{equation}
where $\tau= \frac{2 \pi}{d}m$ is the reciprocal lattice vector.

Using the expression for the Bragg frequency, we can recast
$N^{\parallel}_{ph}$ in the form:

\begin{equation}
\label{eq:number_ph_par1}
 N^{\parallel}_{ph} \approx \alpha Q^2
\frac{1}{4 \sin^2 \theta_B} \frac{1}{k^4_B \Omega^2_2}
\left|
4 \pi \frac{A^{\parallel}_{0}}{1+ i \pi A^{\parallel}_{0}}
\right|^2 k_B L =
\alpha Q^2 \frac{1}{\tau^2 k^2_B \Omega^2_2} \left|
4 \pi \frac{A^{\parallel}_{0}}{1+ i \pi A^{\parallel}_{0}}
\right|^2 k_B L.
\end{equation}

It is also convenient to write (\ref{eq:number_ph_par1}) in the
form (the lattice is assumed to be square and have a period $d$):
\begin{equation}
N_{ph}^{\parallel} \approx \alpha Q^2 \frac{1}{8\pi^3 m^2}
M_{\parallel}
\left(\frac{\lambda_B}{d}\right)^2\frac{L}{\lambda_B}\approx
4\cdot10^{-5}Q^2\frac{1}{m^2}M_{\parallel}
\left(\frac{\lambda_B}{d}\right)^2\frac{L}{\lambda_B},
\label{eq:63a}
\end{equation}
where $M_{\parallel} = \left| 4 \pi \frac{A^{\parallel}_{0}}{1+ i
\pi A^{\parallel}_{0}} \right|^2$.

 Let us analyze the number of quanta (\ref{eq:number_ph_par1})
in the frequency range from 100~GHz to 3 THz.
It will be recalled that the expressions for the coefficients
$\frac{A^{\parallel}_{0}}{1+ i \pi A^{\parallel}_{0}}$, derived in
\cite{nim06},  are valid for the the case when $kR \ll 1$ ($R$ is
the radius of the thread), and so all the following conclusions
are valid for the case $kR \ll 1$.

However, for $kR \geq 1$,  the proposed analysis cannot be
applied, and a more complicated theory is required.

\begin{figure}[h] \epsfxsize = 10 cm
\centerline{\epsfbox{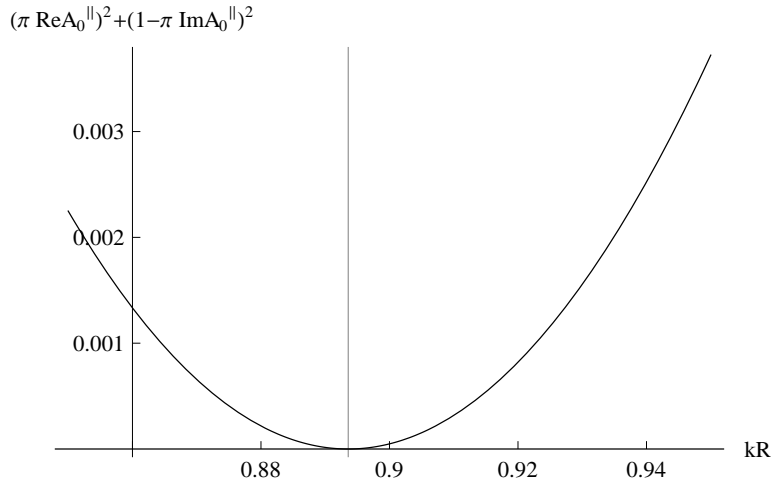}}
\caption{Denominator of $M_{\parallel}$ for perfectly conducting
threads as a function of $kR$}
\label{fig:ZNAMperf2}
\end{figure}

The number of quanta (\ref{eq:63a}) includes the factor
$M_{\parallel}=\left|4\pi \frac{A^{\parallel}_{0}}{1+ i \pi
A^{\parallel}_{0}}\right|^2$, which determines the behavior of the
number of the emitted quanta as a function of the radiation
frequency, the type and the radius of threads.
The denominator of $M_{\parallel}$ has the form $\pi^2
\texttt{Re}\, A0^2 +(1- \pi \texttt{Im}\, A0)^2$ and for
perfectly conducting threads appears to be zero at $kR \approx
0.89$ (see Fig.\ref{fig:ZNAMperf2} and equation (10) in
\cite{nim06}).

So, close to $kR \approx 0.89$,  the factor $M_{\parallel}$ and
therefore the number of quanta exhibit a sharp growth.

\begin{figure}[h] \epsfxsize = 10 cm
\centerline{\epsfbox{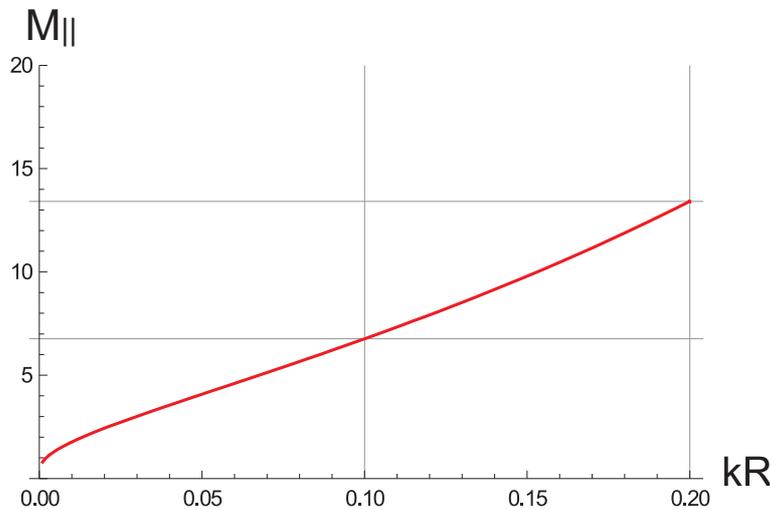}}
\caption{Factor $M_{\parallel}$ for perfectly conducting threads
as a function of $kR$}
\label{fig:MparPERF}
\end{figure}

As follows from Fig.\ref{fig:MparPERF}, in the range $kR\approx
0.2$, the factor $M_{\parallel}$ is of the order of ten.  As a
result, (\ref{eq:63a}) yields the following estimate for the
number of quasi-Cherenkov radiation quanta having a frequency of
about 1THz and  emitted by an electron over 1 cm path length
(it is assumed that $m=1$ and $d\sim 0.3$~cm)
\[
N_{ph}^{\parallel}\approx 10^{-5}M_{\parallel},
\]
for example, for $M_{\parallel}=10$ over the length $L=10$~cm, we
have
\begin{equation}
N_{ph}^{\parallel}=10^{-3}\quad\mbox{quanta}. \label{eq:Nphe}
\end{equation}

Simple estimations can give the number of radiated quanta for an
electron bunch passing through (or moving along the surface) of a
photonic crystal \cite{VG_NO2012,VG_bunch}. Suppose  the bunch
duration is $\tau_b=10^{-12}$~s (i.e., the length of about the
radiation wavelength) and the number of electrons is $n_e = 10^9$
\cite{beam1,beam2}. Then the radiation power, considering quanta
with polarization parallel to the thread's axis, is as follows
\[
P=\frac{N_{ph}^{\parallel} n_e^2 \hbar \omega}{\tau_b} \approx 2
\pi 10^5~W,
\]
and for bunches with the number of electrons  $n_e \sim 10^{11}$
the power approaches a gigawatt level.
In this case, particle bunches obtained at acceleration with
ultra-intense and ultrashort laser pulses are promising for the
creation of a trehertz radiation source with significant power.

Note that (\ref{eq:63a}) includes the dependance on the
uncompensated ion charge $Q$ as $Q^2$, therefore for $Q \sim 30$
we can expect the radiation yield as high as a quantum per
particle under the same conditions as in (\ref{eq:Nphe}).
Thus, photonic crystals can be used to detect high energy charged
particles; moreover, the particle beam need not necessarily pass
through the crystal: it can move along the crystal surface at a
distance $\delta < \frac{\lambda \beta \gamma}{4 \pi}$, which, for
example, for $\lambda \sim 1$~mm and particles with the energy $50
$~MeV ($\gamma=100$) is as large as $\delta \sim 8$~mm.

In conditions of surface diffraction photonic crystals could be
applied for high-energy particle detection by emitting quanta at a
large angle to the particle momentum. Let us note that even in the
optical range, e.g., for $\lambda \sim 10^{-4}$~cm and $\gamma =
10^6 $, the distance $d \sim 10$~cm.

The above-described increase in the number of quanta emitted by a
particle in the range close to $kR\sim 1$ also occurs in the case
of orthogonal polarization. Taking account of a final conductivity
of metals has little influence on the obtained estimate of the
number of quanta. According to  \cite{VG+Zh_Nanophot},  for this
polarization an ordinary Cherenkov effect is possible. As follows
from the above consideration, in the range close to $kR\sim 1$,
the number of Cherenkov radiation quanta also increases. A similar
increase in the number of quanta  is also observed in the case of
diffraction radiation, Smith-Purcell effect, and surface
quasi-Cherenkov radiation, at which particles move along the
surface of such photonic crystals.

\section{Conclusion}
The expressions for spectral-angular distribution of
quasi-Cherenkov radiation emitted by a relativistic particle
traversing a photonic crystal are derived. It is shown that for a
relativistic particle passing through a photonic crystal formed by
periodically strained threads, the intensity of Cherenkov
radiation emitted at small angles to the direction of particle
motion, as contrasted to ordinary Cherenkov radiation, exhibits
anisotropic properties as the photon momentum is rotated about the
direction of particle motion (as the crystal is rotated about the
direction of particle motion at fixed-angle observation of the
outcoming photon).

The intensity of quasi-Cherenkov radiation in terahertz and
optical ranges is shown to be high enough to allow the
experimental study of quasi-Cherenkov radiation.

Some applications of the described phenomena are considered:

In conditions of  surface diffraction photonic crystals could be
applied for high-energy particle detection by emitting a quanta at
a large angle to the particle momentum.

When passing through a photonic crystal, the particle bunches
obtained at acceleration with ultra-intense and ultrashort laser
pulses are promising for the creation of a terahertz  radiation
source with significant power.

\section{Acknowledgement}

Because there is a thermal background in this range,  $\hbar
\omega = k T$, the induced radiation becomes important. We want to
thank Sergei Anishchenko for pointing out that this fact should be
taken into account \cite{ginz,fain,sergei}.

\end{document}